\documentclass[aps,prb,showpacs,twocolumn,superscriptaddress]{revtex4-1}
\usepackage[utf8]{inputenc}
\usepackage[american,british]{babel}
\usepackage[T1]{fontenc}
\usepackage[pdftex]{graphicx}  
\usepackage{xcolor}
\usepackage{dcolumn}
\usepackage{bm}
\usepackage{amsmath,amsthm,amssymb}
\usepackage{verbatim}
\usepackage{ulem}

\definecolor{darkGreen}{RGB}{0,110,0}
\definecolor{darkBlue}{RGB}{0,0,130}
\usepackage[colorlinks,citecolor=darkGreen,linkcolor=darkBlue,urlcolor=blue,hyperindex]{hyperref}

\begin{document}
	
\title{Fractional quantum Hall effect in the interacting Hofstadter model via tensor networks}

\author{M. Gerster}
\affiliation{Institute for Complex Quantum Systems \& Center for Integrated Quantum Science and Technology (IQST), Ulm University, Albert-Einstein-Allee 11, D-89069 Ulm, Germany}

\author{M. Rizzi}
\affiliation{Institut f\"ur Physik, Johannes Gutenberg-Universit\"at Mainz, Staudingerweg 7, D-55099 Mainz, Germany}

\author{P. Silvi}
\affiliation{Institute for Complex Quantum Systems \& Center for Integrated Quantum Science and Technology (IQST), Ulm University, Albert-Einstein-Allee 11, D-89069 Ulm, Germany}
\affiliation{Institute for Theoretical Physics, University of Innsbruck, A-6020 Innsbruck, Austria}

\author{M. Dalmonte}
\affiliation{Abdus Salam International Center for Theoretical Physics, Strada Costiera 11, Trieste, Italy}

\author{S. Montangero}
\affiliation{Institute for Complex Quantum Systems \& Center for Integrated Quantum Science and Technology (IQST), Ulm University, Albert-Einstein-Allee 11, D-89069 Ulm, Germany}
\affiliation{Theoretische Physik, Universit\"at des Saarlandes, D-66123 Saarbr\"ucken, Germany}
\affiliation{Dipartimento di Fisica e Astronomia, Universit\`a degli Studi di Padova, I-35131 Italy}

\date{\today}

\begin{abstract}
	We show via tensor network methods that the Harper-Hofstadter Hamiltonian for hard-core bosons on a square geometry supports a topological phase realizing the $\nu=1/2$ fractional quantum Hall effect on the lattice. We address the robustness of the ground state degeneracy and of the energy gap, measure the many-body Chern number, and characterize the system using Green functions, showing that they decay algebraically at the edges of open geometries, indicating the presence of gapless edge modes. Moreover, we estimate the topological entanglement entropy by taking a combination of lattice bipartitions that reproduces the topological structure of the original proposals by Kitaev and Preskill, and Levin and Wen. The numerical results show that the topological contribution is compatible with the expected value $\gamma = 1/2$. Our results provide extensive evidence that FQH states are within reach of state-of-the-art cold atom experiments.
\end{abstract}

\maketitle

\section{Introduction}
The Harper-Hofstadter model~\cite{hofstadter1976energy} plays an archetypical role in the current understanding of topological quantum matter on a lattice. It encompasses the basic coupling between particles and a background magnetic field, and supports topological bands with finite Chern number for a broad range of fluxes and tunneling rates~\cite{bernevig2013topological}. Those remarkable properties have motivated proposals~\cite{Dalibard2011,*Jaksch2003, *Goldman2014} and recent experiments in both solid-state~\cite{dean2013hofstadter} and cold atom systems~\cite{aidelsburger2013realization,miyake2013realizing,kennedy2015observation,aidelsburger2015measuring,Mancini2015,*Stuhl2015}, which have extensively investigated its non-interacting limit, including the observation of its fractal spectrum~\cite{dean2013hofstadter} and the measurement of a finite Chern number~\cite{aidelsburger2015measuring} in some of its bands. More recently, experiments using bosonic atoms in optical lattices have shown impressive capabilities to approach the {\it strongly interacting} regime~\cite{kennedy2015observation}, with the ultimate goal of stabilizing lattice analogues of fractional quantum Hall (FQH) states~\cite{bernevig2013topological,fradkin2013field}. However, despite promising small system results based on exact diagonalization~\cite{Sorensen2005,Hafezi2007,moller2009composite,sterdyniak2012particle}, in the large flux regime available to experiments -- when the magnetic length is of order of the lattice spacing  -- theoretical evidence of such states at large scales has been lacking, especially regarding smoking guns of topological order -- gapless edge modes, entanglement properties, and many-body Chern numbers.

\begin{figure}
	\includegraphics[width=0.75\columnwidth]{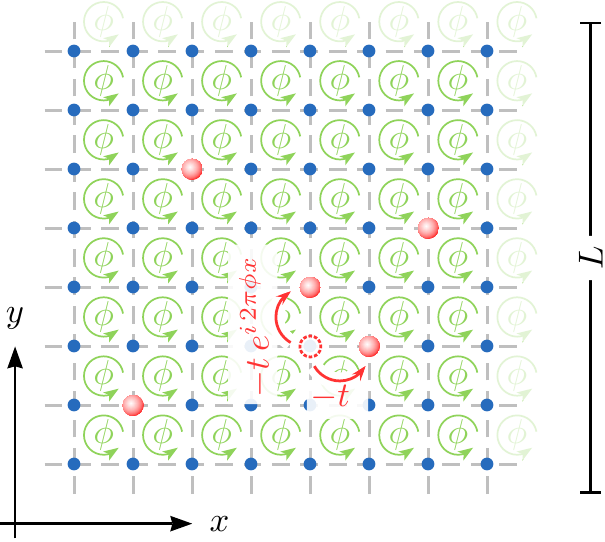}
	\caption{\label{fig:sketch}Sketch of the system: $N$ hard-core bosons (red balls) are hopping on a $L\times L$ lattice, perpendicular to an external magnetic field. The magnetic field gives rise to a flux $\phi$ through each plaquette, which in Landau gauge leads to the indicated phase factors in the hopping amplitudes $t$. The fluxes in the topmost row and rightmost column (faded) are only present in case of periodic boundary conditions (PBC). For PBC, additional phase twists $\theta_x$ and $\theta_y$ can be introduced (see text), allowing for the definition of topological invariants.}
\end{figure}

In this work, we study the strongly-interacting bosonic Hofstadter model, and show that it supports a FQH ground state (GS) akin to the $\nu=1/2$ Laughlin state in the continuum~\cite{fradkin2013field,kalmeyer1987equivalence,regnault2011fractional}. Our analysis is based on a combination of diagnostics, including GS degeneracies on different topologies, the measurement of the many-body Chern number, Green functions' decay at the edge and in the bulk, and a direct measurement of the topological entanglement entropy. The corresponding results serve as a quantitative guideline to address the stability of such phases against temperature: crucially, we show how the spectral gap of the bulk excitations is of the order of $10\%$ of the tunneling rates, showing how FQH states are within reach for temperatures available to current experiments.

The enabling tool of our analysis are numerical simulations based on the tree-tensor network ansatz~\cite{Tagliacozzo2009, Murg2010, *Nakatani2013, Gerster2014}. This class of tailored variational wave functions extends the matrix-product state (MPS) ansatz -- the tensor network class at the heart of the density-matrix renormalization group (DMRG)~\cite{White1992,Schollwock2011}. To obtain the presented results on square lattices with sizes of up to $32 \times 32$ sites, we exploit the reduced scaling of computational costs of the loopless geometry of TTNs, a characteristic not shared by other network structures such as PEPS and MERA~\cite{Verstraete_2006, *Vidal_2007}. Additionally, differently from typical DMRG approaches, TTN warrant direct access to the reduced density matrix of a variety of lattice bipartitions. In the following, we show how this feature enables the direct evaluation of the topological entanglement entropy using a specific combination of various partitions, in parallel to the original proposals of Refs.~\onlinecite{Kitaev2006, *Levin2006}, which are instead not immediately applicable to DMRG studies.

The paper is structured as follows: First, we define the model (Sec.~\ref{sec:model}) and describe the employed numerical method (Sec.~\ref{sec:ttn}). Then, we proceed to present the numerical evidence for a FQH GS in Sec.~\ref{sec:evidence}, based on the low-energy spectrum (Sec.~\ref{sec:en-spectra}), the many-body Chern number (Sec.~\ref{sec:mbcn}), the correlation functions (Sec.~\ref{sec:corr-func}), and the topological entanglement entropy (Sec.~\ref{sec:tee}). In Sec.~\ref{sec:transition} we analyze the robustness of the FQH state upon introducing a superlattice potential. Finally, we conclude our work in Sec.~\ref{sec:conclusion}.

\section{Model Hamiltonian} \label{sec:model}
We study spinless bosonic particles hopping on a $L \times L$ square lattice under the influence of an external magnetic field, as illustrated in Fig.~\ref{fig:sketch}. In the Landau gauge, the system is described by the following Hamiltonian~\cite{Sorensen2005}
\begin{eqnarray}
H &=&U \sum_{x,y} n_{x,y}(n_{x,y} - 1) -t \sum_{x,y} \left\{ a^\dagger_{x+1,y} a_{x,y} \, \mathrm{e}^{-i\, 2\pi \delta_{xL} \theta_x} +\right.\nonumber\\
&+ & \left.a^\dagger_{x,y+1} a_{x,y} \, \mathrm{e}^{i\, 2\pi (\phi x - \delta_{yL} \theta_y)} + \mathrm{h.c.}\right\}
\label{eq:ham}
\end{eqnarray}
of bosonic particles, $[a_{x,y}, a_{x',y'}^{\dagger}] = \delta_{xx'} \delta_{yy'}$, $n_{} = a^{\dagger} a$.
Here, $\phi$ is the magnetic flux through each plaquette (resulting in a magnetic filling factor $\nu=N/(\phi L^2)$, with $N$ the number of bosons in the system) and $\theta_x$, $\theta_y$ implement the twists in the boundary condition~\cite{Hafezi2007}. In the dilute limit (small densities and small fluxes) the lattice physics approaches the one of the continuum~\footnote{Notice that the continuum limit is also recovered by adding tailored long-range hoppings, which effectively flatten the lowest band~\cite{kapit2010exact}.}. However, in the large flux limit, available in cold gases experiments, the phase diagram is not set. On small systems, it has been shown by exact diagonalization (ED) that the GS of the model described by Eq.~\eqref{eq:ham} at filling factor $\nu=1/q$ (where $q$ is an even integer) is compatible with a lattice analogue of the (bosonic) Laughlin wave function~\cite{Hafezi2007,Laughlin1983,moller2009composite,sterdyniak2012particle,Hugel:2016aa}, exhibiting topological GS degeneracy and a non-zero Chern number~\cite{Niu1985, *Tao1986}. However, the overlap with the exact Laughlin wave functions rapidly degrades with system size already for small systems of 6 particles. From a complementary viewpoint, the ladder version of the Hofstadter model has also been shown to share similarities with FQH states~\cite{Budich:2017aa,Strinati:2016aa,Petrescu:2016aa,Haller2017}. Very recently, iDMRG results on cylinders have shown strong signatures of integer quantum Hall states, and have reported fractional current quantization in regimes different from the one we consider here~\cite{He2017}. 
Throughout, we focus on the strongly interacting case $U\rightarrow\infty$ (hard-core bosons) with flux values $\phi=1/8$ and $1/16$ respectively, which correspond to flux setups that are experimentally available. Finally, we fix the energy scale by setting $t=1$.

\begin{figure}
	\includegraphics[width=0.8\columnwidth]{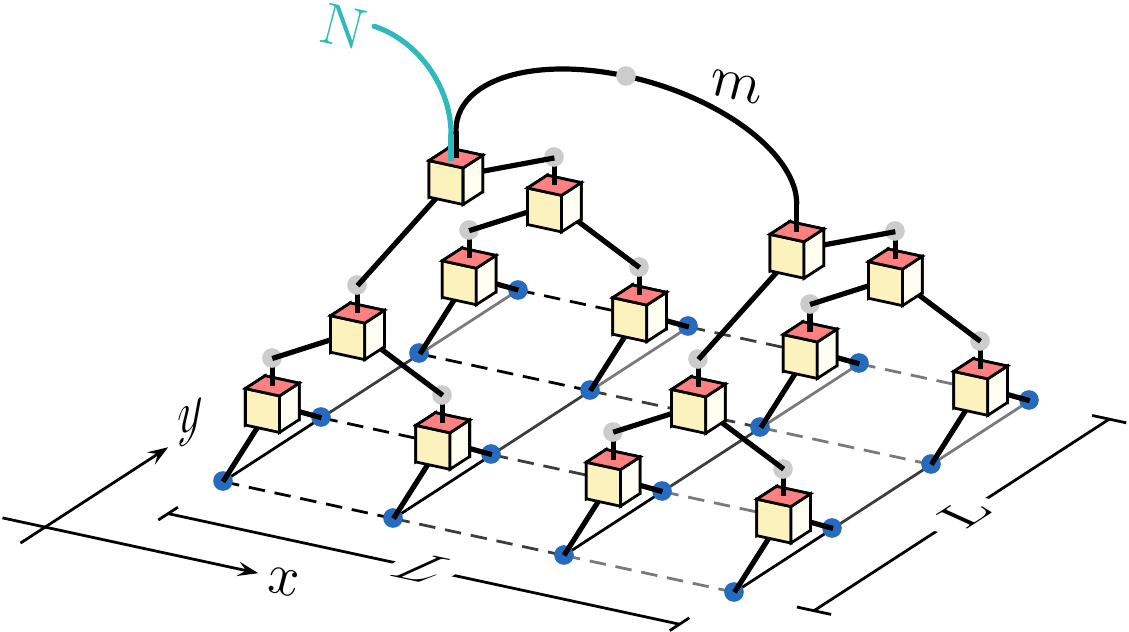}
	\caption{\label{fig:ttn}Binary tree-tensor network ansatz for a $L\times L$ lattice. The blue dots are the physical sites with local dimension~$d$ ($d=2$ for hard-core bosons). Each tensor groups two sites to one virtual site (gray dots), leading to a hierarchical tree structure. In order to capture the 2D lattice geometry, the grouping is performed in $x$- and $y$-direction, alternating from level to level. The dimension of the virtual sites in the~\mbox{$l$-th} level (counting from below) is $\min(d^{2^l}, m)$, where $m$ is the bond dimension of the TTN. The additional cyan link at the top left tensor has dimension one and selects the global particle number $N$~\cite{Singh2011}.}
\end{figure}

\begin{figure*}
	\centering 
	\includegraphics[width=\textwidth]{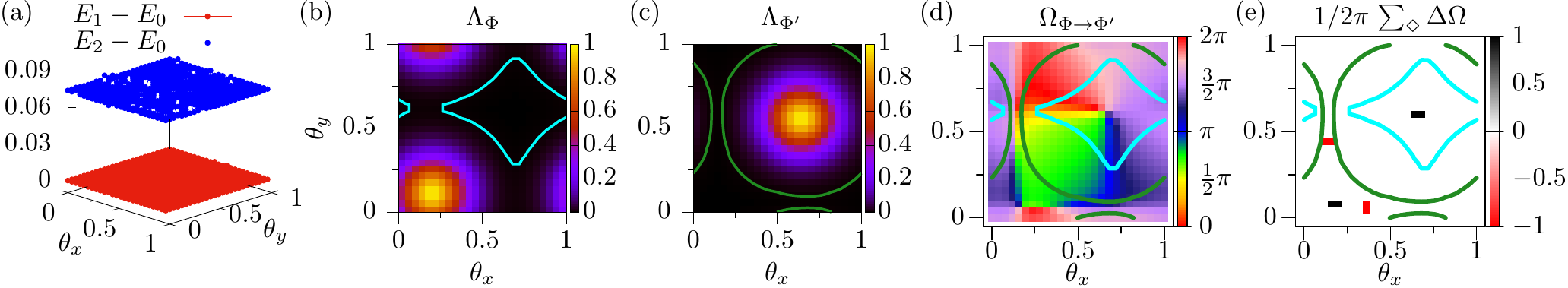}
	\caption{\label{fig:chernnum} Low-lying spectrum (a) and many-body Chern number (MBCN) (b-e) for a system with $L=16$, $N=8$, $\phi=1/16$ ($\nu=1/2$). (a) Finite size spectrum as a function of the twist angles of the boundary conditions. The twofold degenerate GS manifold  is separated from the first excited state by a gap $\Delta\simeq 0.08$. (b-c) Regions where $\Lambda_\Phi$ and $\Lambda_{\Phi^\prime}$, respectively, vanish. The reference multiplet $\Phi$, $\Phi^\prime$ is composed of the two orthogonal GSs at $(\theta_x,\theta_y)=(0.2, 0.12)$, $(0.64, 0.56)$. (d) Color plot of the argument  field $\Omega_{\Phi \rightarrow \Phi^\prime}$ whose count of branch vortices gives the MBCN. (e) Location (and sign) of the branch points, obtained by summing up the angle differences $\Delta\Omega$ along the nearest neighbors of each grid point, resulting in a MBCN of one.}
\end{figure*}

\section{Tree-tensor network ansatz} \label{sec:ttn}
We employ a tree-tensor network (TTN) ansatz~\cite{Shi2006} for the GS and the two lowest excited states to verify the properties discussed above. The specific binary TTN used in this work is illustrated in Fig.~\ref{fig:ttn}, where the standard graphical notation for tensor networks (TN) is employed~\cite{Orus2014, Silvi2017}: tensors are drawn as cubes, with attached lines symbolizing tensor indices (links). Links that are shared by two tensors are contracted, which in TN language means that over their corresponding mutual indices is to be summed. The dimension of each link in the TTN is upper bounded by a constant $m$ (bond dimension), which serves as the refinement parameter of the ansatz: the larger $m$, the more accurate the true many-body state can be approximated. In a binary TTN each tensor has at most three links; therefore, the scaling of the computational resources with the bond dimension is moderate in algorithms using this class of TN states~\cite{Gerster2014}. Furthermore, we exploit particle number conservation by restricting the ansatz to the $N$ particle symmetry sector~\cite{Singh2011}. 

While it is known that a 2D-TTN is not compatible with the area law for the entanglement entropy~\cite{Tagliacozzo2009, Ferris2013}, it possesses several beneficial features which make it a promising tool for the study of intermediate system sizes: (a) The existence of a numerically stable search algorithm for eigenstates~\cite{Murg2010, *Nakatani2013, Gerster2014, Silvi2017}; (b) a low-order polynomial scaling $\mathcal{O}(m^4 L^2)$ of the computational cost; (c) easy interchange of various boundary conditions (open, periodic, twisted); (d) access to the  entanglement entropy for bipartition shapes that enable the determination of the topological entanglement entropy (TEE)~\cite{Kitaev2006, *Levin2006}.
In what follows, we will exploit these properties to gather a number of numerical pieces of evidence supporting a FQH GS of the model Eq.~\eqref{eq:ham} in the case of filling $\nu=1/2$.

\section{Numerical evidence for FQH ground state} \label{sec:evidence}

\subsection{Low-energy spectra} \label{sec:en-spectra}
As first evidence we verify the GS degeneracy, intimately connected to the topological order of the system~\cite{Wen1990,fradkin2013field}. On a torus geometry the GS degeneracy at $\nu=1/2$ filling is expected to be twofold (independent of the twist angles $\theta_x$, $\theta_y$), while the first excited state is to be separated by a bulk gap. We determined the three lowest-energy eigenstates, reported in Fig.~\ref{fig:chernnum}a: We clearly observe a finite energy gap $E_2-E_0\approx 0.1t$ which is typically more than two orders of magnitude larger than the energy difference $E_1-E_0$ between the two states manifold (multiplet) in the GS for the system sizes considered here. Conversely, we verified that for open boundary conditions (OBC) the quasi-degeneracy is removed and we observe that $E_2-E_1 \approx E_1-E_0$.

\subsection{Many-body Chern number} \label{sec:mbcn}
As second evidence we determine the many-body Chern number (MBCN)~\cite{Niu1985, *Tao1986}, which is a direct signature of topological order. For the numerical calculation of the MBCN we follow the prescription put forward by Hatsugai~\cite{Hatsugai_2004, *Hatsugai_2005}, which requires the knowledge of the GS manifold on a 2D grid of twist angles $(\theta_x, \theta_y) \in [0,1]\times [0,1]$ (note that this grid also has a torus geometry). From a practical point of view, this method relies on the calculation of overlaps, a task which can be easily accomplished with TN states: 
More specifically, we need to choose two reference multiplets $\Phi$, $\Phi^\prime$ which have to be non-parallel but otherwise can be arbitrary. (Commonly, one picks two GS multiplets at twist angles far from each other~\cite{Hafezi2007a}.) This corresponds to two different gauge references, which can be used to define two scalar fields: $\Lambda_{\Phi^{(\prime)}}=\det \langle \Phi^{(\prime)}_j | P(\theta_x,\theta_y) | \Phi^{(\prime)}_k \rangle$, where the determinant runs over the indexes $j,k \in \{0,1\}$ of the GS multiplet and 
\begin{eqnarray}
P(\theta_x,\theta_y) &=& | \Psi_0(\theta_x,\theta_y) \rangle \langle \Psi_0  (\theta_x,\theta_y) | +\\
&+& | \Psi_1(\theta_x,\theta_y) \rangle \langle \Psi_1 (\theta_x,\theta_y) |\nonumber
\end{eqnarray}
is the projector on the GS manifold at $(\theta_x,\theta_y)$. If the gauge has been fixed appropriately, $\Lambda_\Phi$ is well-defined (i.e. non-vanishing) where $\Lambda_{\Phi^\prime}$ is not and vice versa, thus forming two complementary regions on the boundary condition torus (an example is shown in Fig.~\ref{fig:chernnum}b-c). Finally, the MBCN is given by the number of branch vortices in the argument field
\begin{equation}
	\Omega_{\Phi\rightarrow\Phi^\prime} = \arg\left( \det \langle \Phi^\prime_j | P(\theta_x,\theta_y) | \Phi_k \rangle \right)
\end{equation}
to be counted (with sign) in any of these two regions~\cite{Hatsugai_2004, *Hatsugai_2005, Hafezi2007a}. We performed this procedure using TTN states; the result is shown in Fig.~\ref{fig:chernnum}d-e. We obtain a MBCN of 1, clearly demonstrating the topological order in the GS of the model. Combined with the two-fold degeneracy discussed above, this shows that the system has a Chern number per state $C = 1/2$, as expected for the $\nu=1/2$ Laughlin state.

\begin{figure}
	\includegraphics[width=\columnwidth]{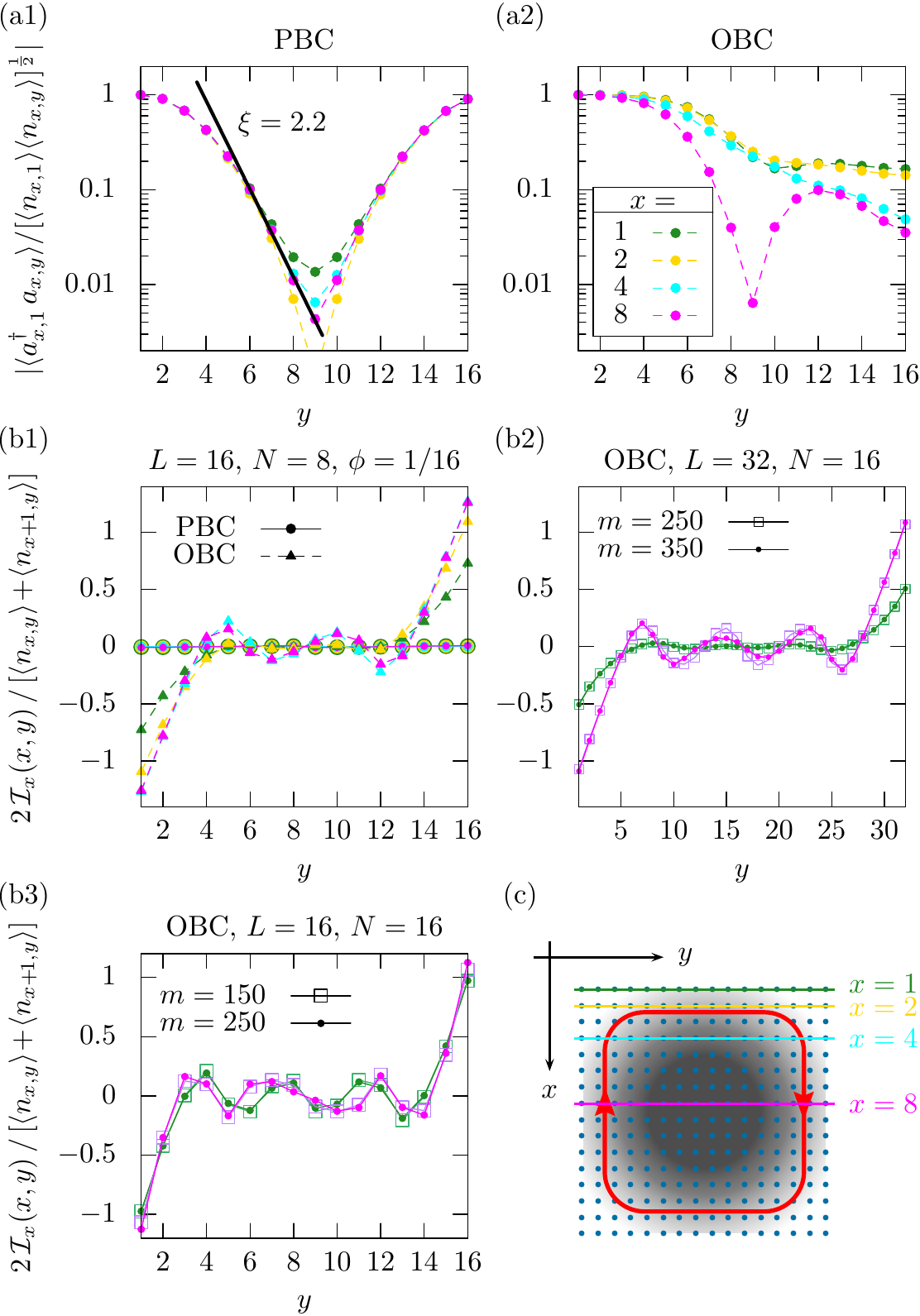}	
	\caption{\label{fig:corr}(a) Green functions along the $y$-direction for PBC (a1) and OBC (a2), for $L=16$, $N=8$, $\phi=1/16$. In the PBC case the decay is exponential, irrespective of the choice of $x$.  In contrast, for OBC, the decay at the edges ($x\approx 1$ or $y\approx L$) is much slower, signaling the presence of edge modes with algebraic correlations. (b) Current in the $x$-direction. (b1): comparison between PBC (no current) and OBC (currents at the edges). (b2): OBC current for $L=32$, $\phi=1/32$, and (b3): for $\phi=1/8$, $L=16$, each with two different bond dimensions~$m$. All panels use the same color code for $x$, which is illustrated in (c). The cartoon in (c) shows the edge current (red arrow) and demonstrates why $\mathcal{I}_x$ vanishes in the bulk (dark-shaded background), while at the edges (bright background) it is nonzero whenever $y \approx 1$ or $y \approx L$. Moreover it is obvious that the sign of $\mathcal{I}_x$ at $y \approx 1$ is opposite to the sign of $\mathcal{I}_x$ at $y \approx L$.}
\end{figure}
	
\subsection{Correlation functions and edge modes} \label{sec:corr-func}
Indirect evidence of a fractional quantum Hall state can also be gathered by monitoring the behavior of the Green function in the system $\mathcal{G}(x,x';y,y') = \langle a^\dagger_{x,y}a_{x',y'}\rangle$. The Green function in periodic systems contains information about bulk excitations: in our case, it is expected to decay exponentially as a function of distance. In contrast, for open boundaries the Green function is expected to reveal the existence of gapless edge modes. Indeed, in the case of $\nu=1/2$, the phase field operator of the corresponding chiral Luttinger liquid edge mode is expected to have considerable overlap with the creation and annihilation operators on the lattice. 
In Fig.~\ref{fig:corr}a, we plot the normalized $\mathcal{G}(x,x;1,y)$ as a function of $y$ both for PBC and OBC. In the case of PBC, the results clearly show that $\mathcal{G}$ decays exponentially as a function of distance, with a correlation length of approximately 2 sites - and thus, considerably smaller than our system sizes. Since for OBC  translational invariance is broken, we consider different values of $x$ (marked by different colors as also illustrated in Fig.~\ref{fig:corr}c).  Here, two distinct regimes are visible: along the edge (here for $x\lesssim 4$) the correlation decays as a power law, indicating a gapless mode localized close to the boundary. In sharp contrast, far from the edges ($x \simeq L/2$) the decay becomes exponential, consistent with the PBC results. Our results on the correlation functions thus strongly confirm a finite bulk gap, coexisting with gapless edge modes. 

An alternative route to detect chiral edge modes, likely more accessible in experiments, is given by the particle current density. Specifically, we consider here the current in the $x$-direction 
\begin{equation}
\mathcal{I}_x(x,y) = i\, \langle a^\dagger_{x+1,y} a_{x,y} - a_{x+1,y} a^\dag_{x,y} \rangle,
\end{equation} 
again both for PBC and OBC. In Fig.~\ref{fig:corr}b we show the normalized $\mathcal{I}_x(x,y)$ as a function of $y$ and for different values of $x$ (again with the same color code). The results are in strong agreement with the ones obtained from the Green function analysis: For PBC, where there is only bulk, the current vanishes throughout the system, while for OBC we observe strongly enhanced currents along the edges, i.e. at $y\approx1$ or $y\approx L$. In Fig.~\ref{fig:corr}c we provide an illustration of the edge current present in an OBC lattice, giving an intuitive (albeit over-simplified) explanation for the behavior of the currents plotted in Fig.~\ref{fig:corr}b.

\begin{figure}
	\includegraphics[width=0.9\columnwidth]{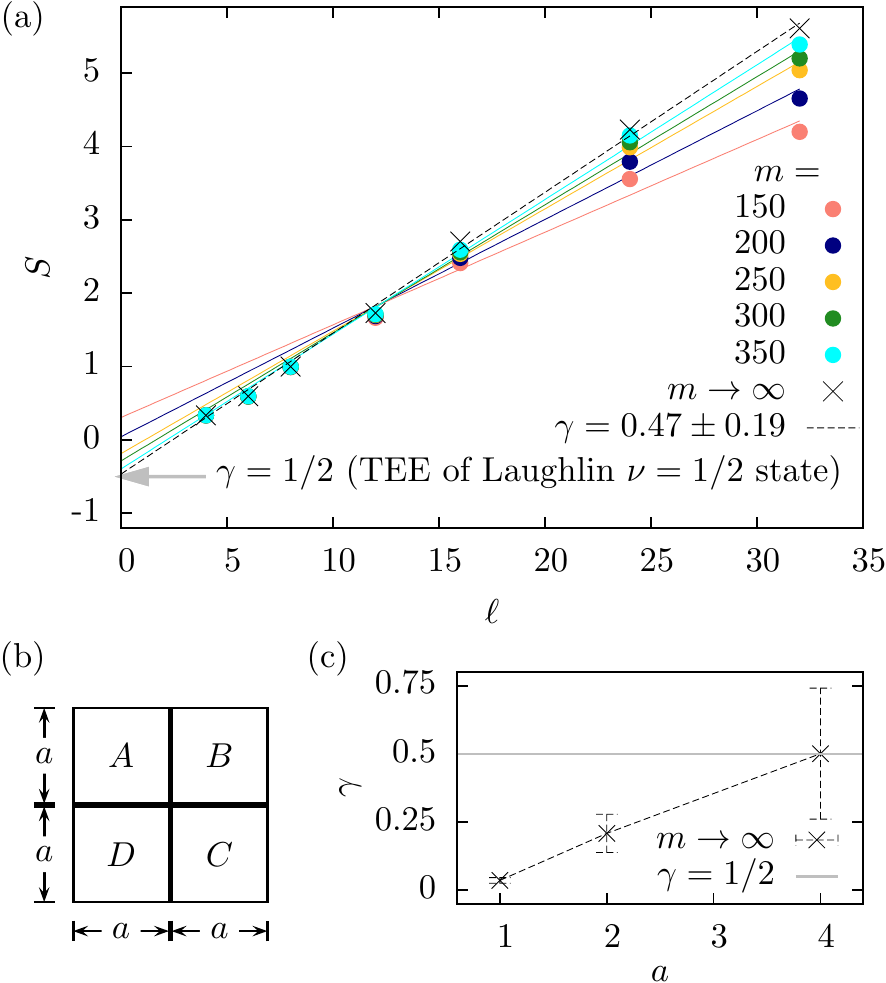}
	\caption{\label{fig:entropy}(a) TEE $\gamma$ obtained from an extrapolation in $m$: Lines are linear fits through data points with $\ell>10$, for $L=16$, $N=16$, $\phi=1/8$. The resulting $\gamma$ is compatible with $1/2$. (b) Arrangement of square partitions for the extraction of $\gamma$~\cite{Kitaev2006,*Levin2006} and resulting TEE (c) as a function of the block size $a$. For large enough $a$, the data is compatible with $\gamma=1/2$.}
\end{figure}

\subsection{Topological entanglement entropy} \label{sec:tee}
An unambiguous indicator for topological order is the topological entanglement entropy (TEE) $\gamma$, as introduced in Refs.~\onlinecite{Kitaev2006, *Levin2006}. A finite value of $\gamma$ signals the presence of anyonic excitations above the GS manifold - remarkably, without having to directly access excited states. In particular, $\gamma$ is directly related to the quantum dimensions of the excitations.
Extracting the TEE in numerical simulations is however challenging. System sizes available to exact diagonalization are typically too small to measure $\gamma$ in a meaningful fashion. DMRG simulations can reach considerable system sizes: in that context, one tries to extract the TEE $\gamma$ as a correction to the perimeter-law scaling of the entanglement entropy: $S(\ell) = c \, \ell - \gamma$, where $S(\ell)$ is the von Neumann entropy $S = - \mathrm{tr}[ \rho \log_2 \rho]$ for a spatial partition of perimeter $\ell$.
In Fig.~\ref{fig:entropy}a, we show our results for $S(\ell)$ for PBC with $L=16$. In order to avoid strong finite size effects, we excluded from our fits the data with $\ell< 10$. After extrapolation to $m\rightarrow\infty$, we get $\gamma=0.47\pm0.19$, which is close to the value obtained with our maximum bond dimension. Despite the large error bar, mostly due to the fact that we have only access to few points in $\ell$, our result show that the system has $\gamma>0$, and the result is compatible with the expected TEE for a Laughlin state with $\nu=1/q$, $\gamma=\log_2 \sqrt{q}$. 
While this procedure has been well tested in several models~\cite{depenbrock2012nature,Jiang_2012}, it is desirable to apply a method for obtaining the TEE which follows directly the original prescription: Such technique operates on regions with different shapes in order to exclude spurious effects. Within our TN ansatz this is possible considering the geometry in Fig.~\ref{fig:entropy}b, where it can be shown that the following relation holds:
\begin{eqnarray}
	-\gamma &=& S_A + S_B + S_C + S_D + \\
	&-& S_{AB} - S_{BC} - S_{CD} - S_{DA} + S_{ABCD} \; . \nonumber
\end{eqnarray}
The basic building block of this procedure are square partitions of size $a\times a$. The results of the corresponding simulations, extrapolated in $m\rightarrow\infty$, are shown in Fig.~\ref{fig:entropy}c: for $a=4$ the results indicate a finite TEE of $\gamma\simeq 0.5$, again with rather large error bars (owing to the uncertainty in the $m$ extrapolation), but otherwise in good agreement with the expected behavior for $\nu=1/2$ FQH states. We note that for $a=1,2$ the results are far from this prediction, as expected, since for those values the partition size is smaller than the correlation length $\xi$ as computed from the Green function decay.

\begin{figure}
	\includegraphics[width=1\columnwidth]{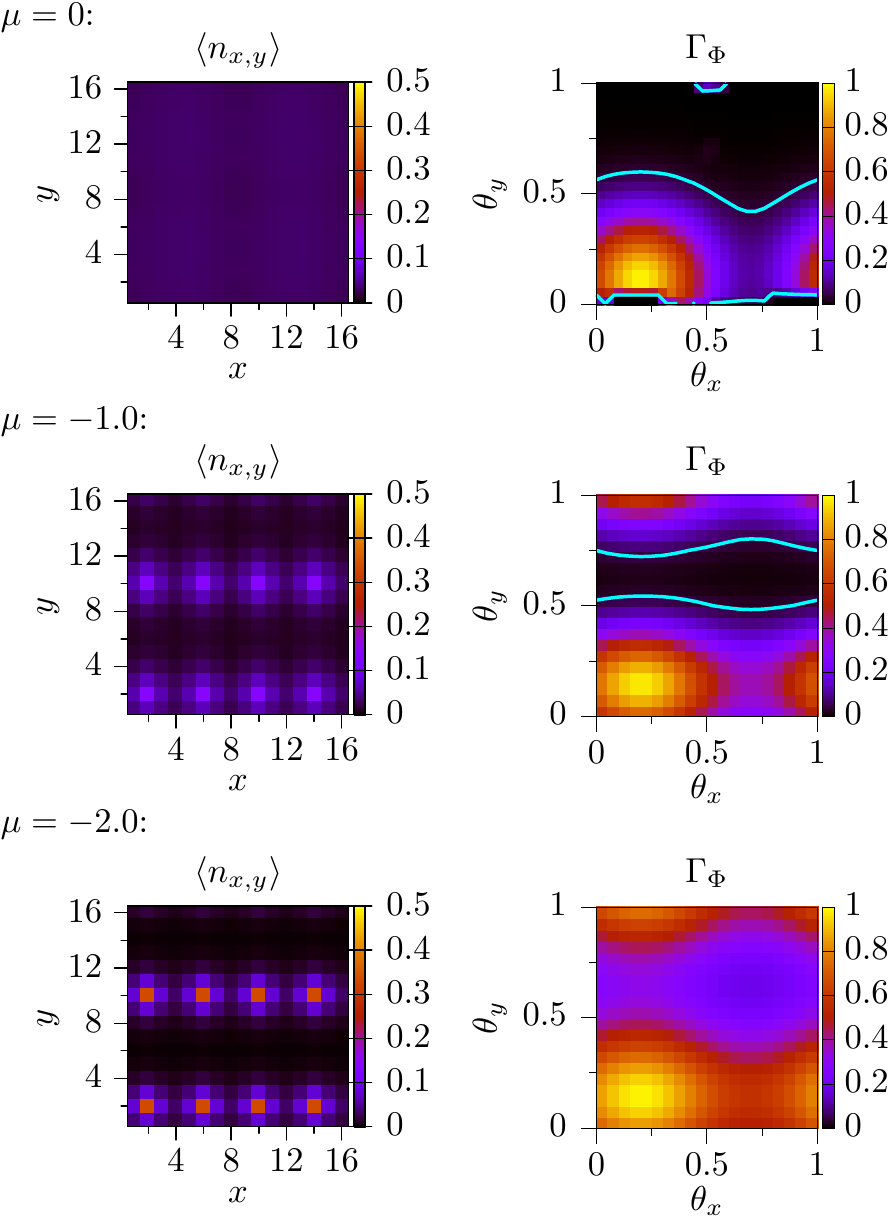}
	\vspace{-6mm}
	\caption{\label{fig:suplatt} Left column: on-site occupations $\langle n_{x,y} \rangle$ of the lattice sites for different values of $\mu$, showing how the particles gradually localize at the energetically favored sites of the superlattice. Right column: plots of $\Gamma_\Phi$ for different values of $\mu$. For $\mu \gtrsim -1.0$ there exist regions (visible as black areas enclosed by cyan lines) where $\Gamma_\Phi$ vanishes, indicating topologically non-trivial character. System parameters are $L=16$, $N=8$, $\phi=1/16$.}
\end{figure}

\section{Phase transition from FQH state to trivial insulator} \label{sec:transition}
\begin{figure}
	\includegraphics[width=0.9\columnwidth]{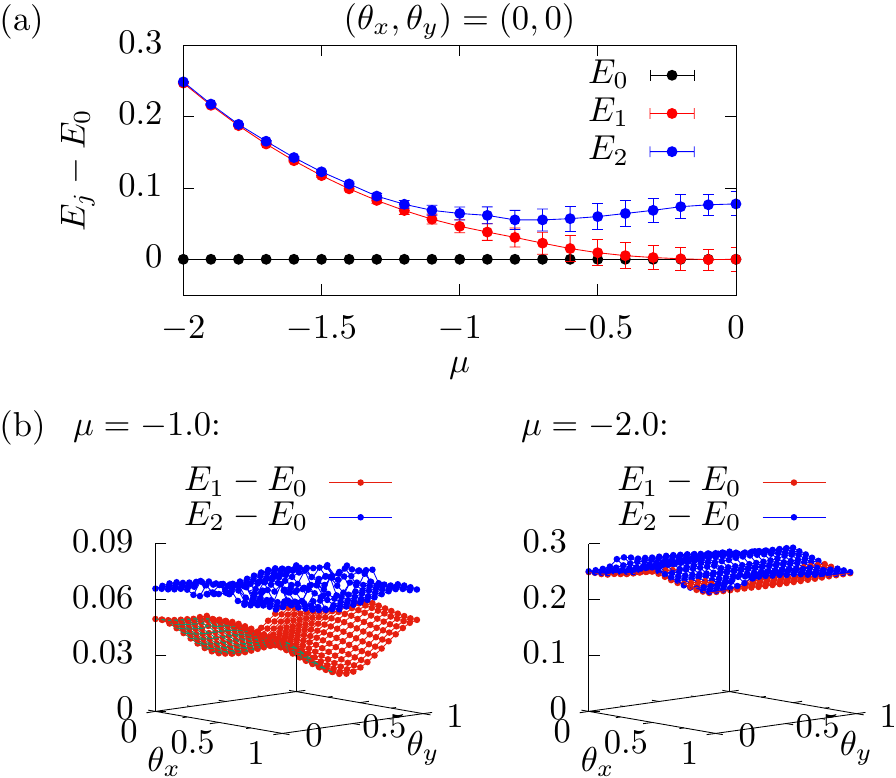}
	\caption{\label{fig:mu_gaps}(a) Three lowest eigenenergies of $\tilde{H}$ as a function of~$\mu$, at $\theta_x=\theta_y=0$. (b) Low-energy spectra on the whole boundary condition torus for $\mu=-1.0$ and $\mu=-2.0$. System parameters are $L=16$, $N=8$, $\phi=1/16$.}
\end{figure}

\begin{figure}
	\includegraphics[width=0.95\columnwidth]{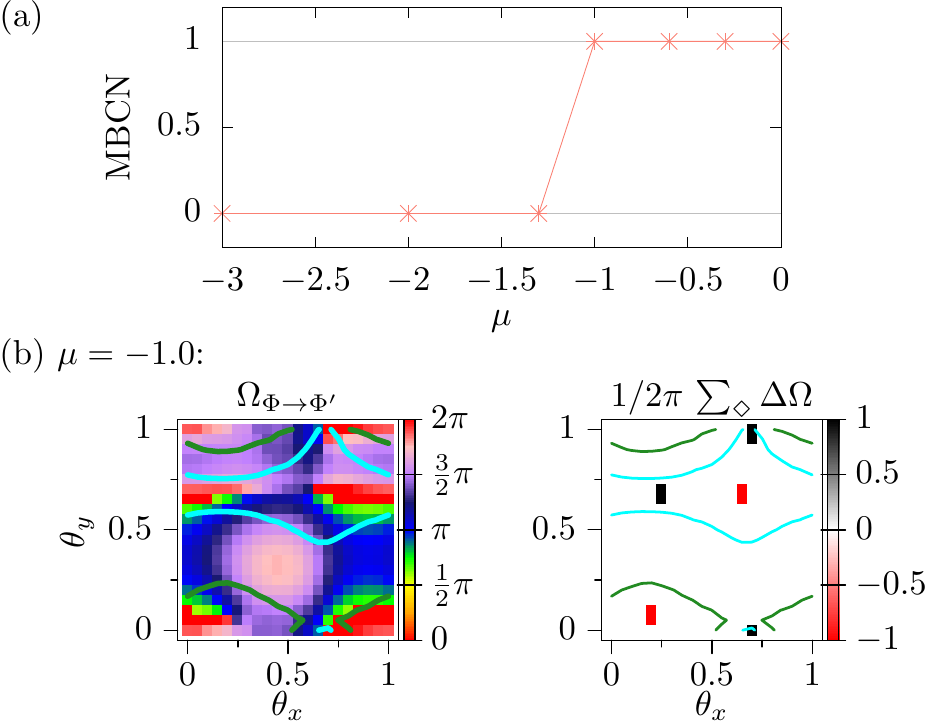}
	\caption{\label{fig:mu_mbcn}(a) MBCN as a function of $\mu$, displaying a transition at $\mu_c \approx -1$. (b) Determination of MBCN for $\mu=-1.0$, resulting in $\mathrm{MBCN}=1$. System parameters are $L=16$, $N=8$, $\phi=1/16$.}
\end{figure}

Our diagnostics also allows for the detection of possible phase transitions. Here we demonstrate this by investigating the robustness of the FQH state upon introducing a superlattice potential into the model from Eq.~\eqref{eq:ham}, which, in turn, can be engineered in cold atoms experiments~\cite{Sebby-Strabley2006, *Foelling2007}. We consider the Hamiltonian
\begin{equation}
\tilde{H} = H + \mu \sum_{(x,y) \in \, \mathrm{sup.latt.}} n_{x,y} \; ,
\label{eq:ham_suplatt}
\end{equation}
where $\mu \leq 0$ denotes the depth of the superlattice potential. We focus on the case where the number of sites of the superlattice is equal to the number of bosons $N$ in the system. For sufficiently large $|\mu|$, the GS of $\tilde{H}$ is a product state with the particles pinned at the superlattice minima (see Fig.~\ref{fig:suplatt}) which is, by construction, topologically trivial. The transition between the two qualitatively different GSs occurs at some critical potential depth $\mu_c < 0$. In order to detect this transition we use the energy spectra on the boundary condition torus, and the MBCN, as shown in Figs.~\ref{fig:mu_gaps} and~\ref{fig:mu_mbcn}. With increasing depth of the potential, we observe how the GS quasi-degeneracy is removed (see Fig.~\ref{fig:mu_gaps}).
As long as the GS multiplet is still quasi-degenerate, i.e. separated from the first excited state ($E_2 - E_1 \gtrsim E_1 - E_0$) on the whole torus of twist angles, we measure a MBCN of one (see Fig.~\ref{fig:mu_mbcn}). 
At the critical point $-1.3 \lesssim \mu_c \lesssim -1.0$ this condition is no longer met and the MBCN vanishes, signaling a disappearance of the topological order. 
The topologically trivial character of such a GS can be evidenced by considering the quantity 
\begin{equation}
\Gamma_\Phi(\theta_x, \theta_y) = \langle \Phi | \Psi_0(\theta_x, \theta_y) \rangle \langle \Psi_0(\theta_x, \theta_y) | \Phi \rangle \; , 
\end{equation}
where $\Phi$ represents a gauge reference, formed by a single GS at an (arbitrary) reference point $(\theta^{[r]}_x, \theta^{[r]}_y)$. If the GS of the system is in fact topologically trivial, $\Gamma_\Phi$ is well-defined (i.e. non-vanishing) on the whole boundary condition torus. On the contrary, this is not the case if the GS has non-trivial topological character: in that case there is a finite region on the torus where $\Gamma_\Phi$ vanishes. Whenever this happens, the GS is quasi-degenerate and Hatsugai's method for determining the MBCN (as described in Sec.~\ref{sec:mbcn}) is applicable. In Fig.~\ref{fig:suplatt} we show $\Gamma_\Phi$ for different values of $\mu$, demonstrating how it gradually becomes well-defined on the whole boundary condition torus as the depth of the superlattice potential is increased.

\section{Conclusions and outlook} \label{sec:conclusion}
We have presented extensive numerical evidence supporting the existence of a fractional quantum Hall phase in the Harper-Hofstadter Hamiltonian for hard-core bosons on a square lattice. Our analysis considered a wide range of independent diagnostics, including spectral properties, the many-body Chern number, correlation functions, currents, and entanglement entropies, and shows how TTN algorithms provide a flexible tool to address the interplay of gauge fields and interactions in two-dimensional systems. The results indicate that the correlation length is typically of the order of a few lattice sites, with corresponding gaps of order $0.1t$: these signatures point toward the fact that lattice fractional quantum Hall states can be realized in present cold atom experiments, as well as potentially in cavity array experiments~\cite{Kapit:2014aa,Owens:2017aa}. The temperatures and sizes required match those already available in experiments, and for which adiabatic state preparation protocols have very recently been shown to be applicable~\cite{He2017,Motruk:2017aa}.

\begin{acknowledgments}
We thank M. Burrello and A. Sterdyniak for a careful reading of the manuscript, J. Budich and H.-H. Tu for discussions, and F. Tschirsich for contributing numerical libraries. Numerical
calculations have been performed with the computational resources provided by the
bwUniCluster project~\footnote{{b}wUniCluster: funded by the Ministry of Science, Research and Arts and the universities of the state of Baden-W{\"u}rttemberg,  Germany, within the framework program bwHPC.}, the JUSTUS project, CINECA via the TEDDI project, and the MOGON cluster at the JGU Mainz. We acknowledge financial support from EU projects RYSQ and UQUAM, the German Research Foundation (DFG) through the SFB/TRR21, OSCAR and TWITTER, and the Baden-W\"urttemberg Stiftung via Eliteprogramm for postdocs. S.M. gratefully acknowledges the support of the DFG via a Heisenberg fellowship.
\end{acknowledgments}

\appendix

\begin{figure*}
	\includegraphics[width=\textwidth]{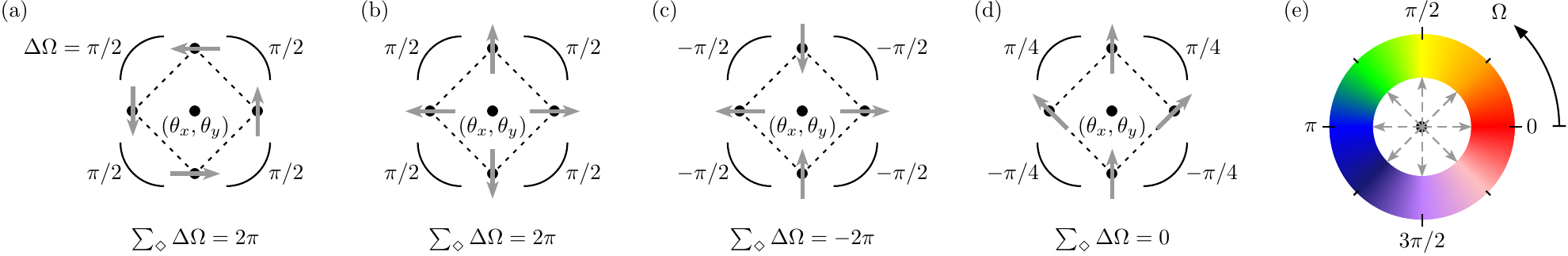}
	\caption{\label{fig:vortices}(a-d) Four example configurations for the argument field $\Omega_{\Phi\rightarrow\Phi^\prime}$ around a point $(\theta_x, \theta_y)$ on the boundary condition grid. The branch vortex count $1/2\pi \sum_\diamond \Delta\Omega$ is $1$ for the configurations shown in (a) and (b), $-1$ for the configuration in (c), while for the one in (d) it is $0$. (e) Illustration of the color code for $\Omega_{\Phi\rightarrow\Phi^\prime}$, used in Figs.~\ref{fig:chernnum}d and~\ref{fig:mu_mbcn}b.}
\end{figure*}

\section{Algorithm for ground and low excited states with tree-tensor networks}
The ground state search is performed according to a variational scheme, first developed in the context of DMRG/MPS ground state algorithms~\cite{Schollwock2011}, and which in the meantime has been generalized to arbitrary loopless TN geometries~\cite{Murg2010, *Nakatani2013, Gerster2014, Silvi2017}. The basic motivation is the following: Although the TN representation of the many-body state already reduces dramatically the number of coefficients in the state vector (as compared to the full Hilbert space dimension), solving the minimization problem
\begin{equation}
E_0 = \langle \Psi_0 | H | \Psi_0 \rangle \stackrel{!}{=} \mathrm{min.}
\label{eq:gs_energy}
\end{equation}
for the entire TN state $|\Psi_0\rangle$ directly is in practice not feasible, because the number of coefficients in $|\Psi_0\rangle$ is typically still far too large to be handled by numerical eigensolvers. Instead, one applies an iterative strategy: out of all the tensors, which together constitute the TN state, only the coefficients of one (or two, dependent on the update scheme~\cite{Hubig2015,Silvi2017}) tensor(s) are considered variational, while the others are taken to be fixed. This allows one to contract the physical Hamiltonian $H$ to a reduced, effective Hamiltonian $H_\mathrm{eff}$, only acting on the degrees of freedom of the variational tensor(s). For loopless TNs the resulting reduced optimization problem can again be formulated as a standard eigenvalue problem, whose size is now manageable by an eigensolver. One then successively targets all the tensors in the TN, thereby gradually decreasing the energy expectation value $\langle \Psi_0 | H | \Psi_0 \rangle$. This procedure is called {\it sweeping}. A sweep is completed after all the tensors in the TN have been updated once. For the TTN architecture employed here (sketched in Fig.~\ref{fig:ttn}), the ansatz contains $N_s-2$ tensors ($N_s=L^2$: number of lattice sites) and all involved contractions (both computation of $H_\mathrm{eff}$ and solving the reduced eigenvalue problem) have computational complexity $\mathcal{O}(m^4)$ or less. Therefore, the GS algorithm runtime scales as $\mathcal{O}(N_s \, m^4)$. Convergence of the GS energy is typically reached after less than ten sweeps.

To determine excited states, we employ a very similar algorithm to the one just described, with the small modification that orthogonality to all previously determined eigenstates is enforced. This can be achieved by penalizing overlap with these eigenstates; more in detail, in order to obtain the $n$-th excited state of $H$ (orthogonal to all lower-lying eigenstates $|\Psi_k \rangle$, $k \in [0, \, n-1]$) we solve the optimization problem
\begin{equation}
E_n = \langle \Psi_n | H | \Psi_n \rangle + \sum_{k=0}^{n-1} \epsilon_k \langle \Psi_n | P_k | \Psi_n \rangle \stackrel{!}{=} \mathrm{min.} \; ,
\label{eq:ex_energy}
\end{equation}
where $P_k=|\Psi_k \rangle \langle \Psi_k |$ is the projector on the $k$-th eigenstate and $\epsilon_k$ is as an energy penalty which has to be chosen large enough, which means larger than the energy difference $|E_k - E_n|$ to the target state. Of course this energy difference is not known at the start of the algorithm, but in practice one can simply estimate a value which is guaranteed to be large enough, e.g. one can set $\epsilon_k$ to be one order of magnitude larger than a typical energy scale in the system. The scaling of the computational complexity of the algorithm is not changed by the additional projective terms in Eq.~\eqref{eq:ex_energy}: In complete analogy to the effective Hamiltonian these terms lead to effective projectors, which only contribute a (typically small) overhead to the algorithm. In particular, the runtime scaling for determining excited states remains at $\mathcal{O}(N_s \, m^4)$.

\begin{figure}
	\includegraphics[width=\columnwidth]{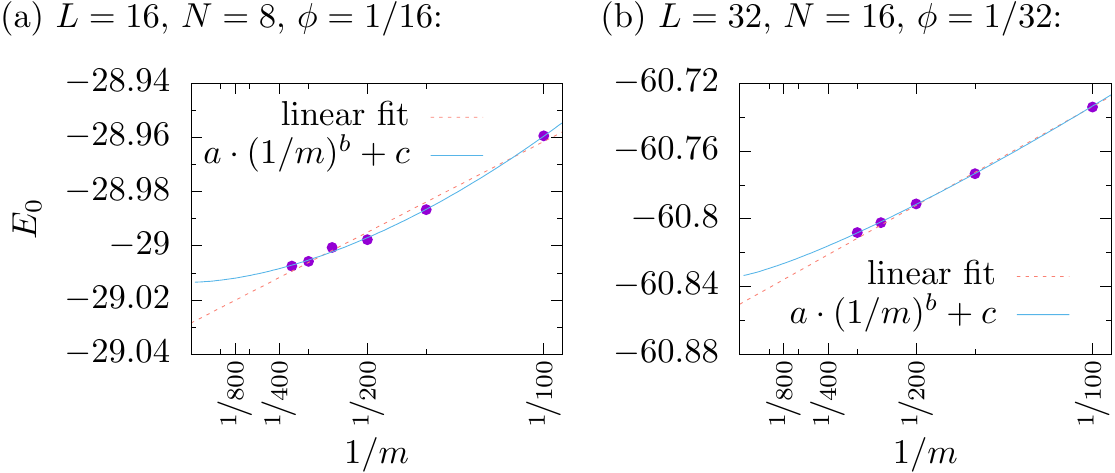}
	\caption{\label{fig:m_energy}GS energy $E_0$ as a function of bond dimension $m$ for two different system sizes $L=16$ (a) and $L=32$ (b). Also shown are two different $m \rightarrow \infty$ extrapolations, a simple one linear in $1/m$ and a more versatile one where the exponent of $1/m$ is allowed to be adjusted by the fit. This procedure can be used to estimate the ground state energies and corresponding errors to be $E_0 = -29.015 \pm 0.016 $ (a), and $E_0= -60.84 \pm 0.03$ (b).}
\end{figure}
\begin{figure}
	\includegraphics[width=\columnwidth]{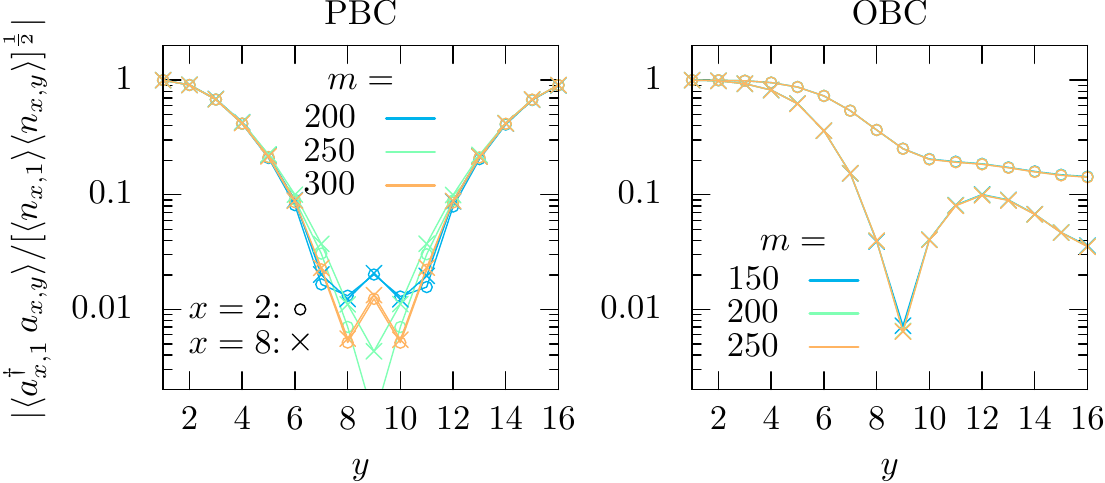}
	\caption{\label{fig:m_corr}Bond dimension convergence of the Green functions from Fig.~\ref{fig:corr}a. Different colors denote different bond dimensions, while the two different symbols refer to two different choices of $x$ (circles: $x=2$, crosses: $x=8$). For PBC the different choices of $x$ result in the same curves, as expected for a system with translational invariance; only at $y\approx L/2$ translational invariance is slightly broken, indicating a finite bond dimension error of the order of $10^{-2}$.}
\end{figure}

\section{Procedure for determining the location of branch vortices}
As described in Sec.~\ref{sec:mbcn}, we obtain the branch vortex count of a point $(\theta_x, \theta_y)$ on the boundary condition grid by summing up the angle differences $\Delta \Omega$ of the argument field $\Omega_{\Phi\rightarrow\Phi^\prime}$ along the four nearest neighbors of $(\theta_x, \theta_y)$. This procedure, following Refs.~\onlinecite{Hatsugai_2004, *Hatsugai_2005}, is exemplified in Fig.~\ref{fig:vortices} for four different illustrative configurations of $\Omega_{\Phi\rightarrow\Phi^\prime}$, displaying different branch vorticities.  Moreover, in Fig.~\ref{fig:vortices}e we provide an illustration for the color code used in Figs.~\ref{fig:chernnum}d and~\ref{fig:mu_mbcn}b.

\section{Bond dimension convergence of observables}
We exemplify the convergence of the energy expectation value as a function of the bond dimension $m$ in Fig.~\ref{fig:m_energy}. Estimates for finite bond dimension errors can be obtained by extrapolating to the limit $m\rightarrow \infty$. For the system sizes considered here, we reached bond dimensions of up to $m\approx 500$, typically displaying truncation errors of order of $10^{-6}$. In Fig.~\ref{fig:m_corr} we demonstrate bond dimension convergence for the Green functions shown in Fig.~\ref{fig:corr}a.

\bibliography{references}

\end{document}